\documentclass[showpacs,preprint,aps]{revtex4}
\usepackage{graphicx}
\usepackage{times}

\begin{document}

\title{Hydrogen induced surface metallization of $\beta$-SiC(100)-($3\times 2$)
revisited by DFT calculations }
\author{R. Di Felice\footnote{rosa@unimore.it}, C. M. Bertoni, C. A. Pignedoli\footnote{Present
Address: IBM Corporation, Zurich Research Laboratory,
Saumerstrasse 4, Rueschlikon, Switzerland}}
\affiliation{INFM-National Research Center on nanoStructures and
bioSystems at Surfaces (S$^{3}$), Dipartimento di Fisica,
Universit\`a di Modena e Reggio Emilia, Via Campi 213/A, 41100
Modena, Italy}
\author{A. Catellani}
\affiliation{CNR-IMEM, Parco Area delle Scienze 37a, 43010 Parma, 
Italy, and INFM-S3}

\begin{abstract}
Recent experiments on the silicon terminated $3\times 2$ SiC(100)
surface indicated an unexpected metallic character upon hydrogen
adsorption. This effect was attributed to the bonding of hydrogen
to a row of Si atoms and to the stabilization of a neighboring
dangling bond row. Here, on the basis of Density-Functional
calculations, we show that multiple-layer adsorption of H at the
reconstructed surface is compatible with a different geometry:
besides saturating the topmost Si dangling bonds, H atoms are
adsorbed at rather unusual sites, \textit{i.e.} stable bridge
positions above third-layer Si dimers. The results thus suggest an
alternative interpretation for the electronic structure of the
metallic surface.
\end{abstract}

\date{\today}
\maketitle



The silicon terminated $\beta$-SiC(100)-$3\times 2$ surface is
constituted of extra Si-dimers on top of silicon terminated layers
\cite{vic-rev,cat-rev,pollmann99}. A recent experimental
investigation \cite{bermudez,souk,amy03} has shown a surprising
property: the transition from a semiconducting to a metallic
surface induced by hydrogen chemisorption. In general
semiconductor surfaces undergo reconstruction, relaxation or
dimerization just to eliminate the presence of partially occupied
surface bands within the gap. Furthermore, hydrogen chemisorption
on the reconstructed surfaces can modify or even quench the
reconstruction by passivating the surface, often removing all the
surface states from the gap. For this system, instead, the
experiments have been interpreted in terms of an opposite effect,
\textit{i.e.} the existence of a band of unsaturated dangling
bonds (DBs) on a row of Si atoms, induced by the adsorption of H
atoms along a neighboring row of other Si atoms. On the basis of
their observations, Derycke et al. \cite{souk} pointed out the
possibility of engineering, at the atomic level, the structural
and electric properties of surfaces using the reconstruction
geometry as a template \cite{souk}.

The interest in this silicon terminated, overstoichiometric
$\beta$-SiC(100) surface is linked to the problem of the epitaxial
production of silicon carbide, with a high technological impact
\cite{souk96}. The presence of excess Si atoms allows the system
to stabilize the subsurface planes by the creation of multiple
Si-dimer layers that are responsible for different possible
reconstructions. Among them, the observed $3\times 2$ periodicity
is obtained with a complex layering of dimers of different
lengths. The stable geometry of this reconstruction is still
intensely debated: here, we do not enter this controversy, but
rather focus on the mechanisms of H adsorption, assuming for the
clean surface the model proposed by Lu et al. \cite{pollmann99},
consistently with the choice adopted for the interpretation of the
experimental H-induced metallization \cite{souk}.
\begin{figure}[tbp]
\includegraphics[scale=0.4]{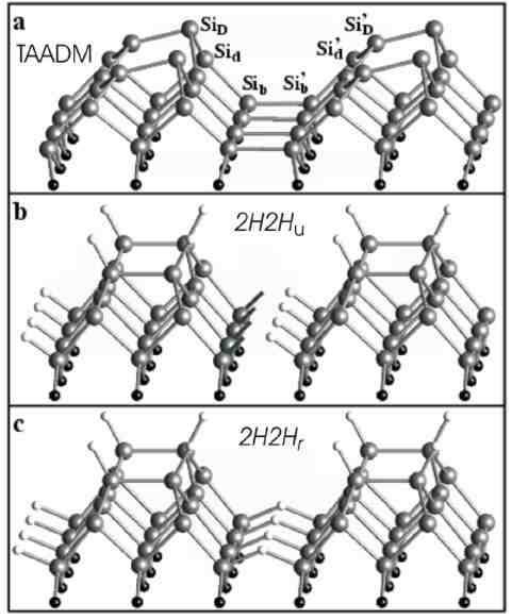} 
\caption{Side view of the outermost layers for the 3$\times$2
model reconstruction adopted in the present work: black (gray)
spheres represent C (Si) atoms, smaller white spheres indicate H
atoms. (a) The clean reconstructed TAADM SiC(100) surface. (b) The
\textsl{2H2H}$_u$ model for the hydrogenated surface, proposed on
the basis of the recent experiments \cite{souk,amy03}. (c) The
equilibrium adsorption geometry \textsl{2H2H}$_r$ obtained by
relaxing the system starting from the experimental
\textsl{2H2H}$_u$ model. A 2$\times$2 replica of the unit surface
cell is shown.}
\end{figure}
As indicated in Fig.~1a, this clean $3\times 2$ surface is
characterized by a coverage of 1 monolayer (ML) extra silicon
atoms on top of the bulk-like Si-terminated SiC(100) face. Of this
extra ML, $\frac{2}{3}$ ML Si atoms are accomodated at the first
adlayer forming a couple of flat dimers in each periodicity unit.
The remaining $\frac{1}{3}$ ML Si atoms are accomodated on top of
them forming a second adlayer of buckled dimers. The Si-rich
SiC(100)-$(3\times 2)$ surface thus exposes 3 outermost Si layers,
starting from the vacuum, with coverages of $\frac{1}{3}$ ML,
$\frac{2}{3}$ ML, and 1 ML on the first, second, and third layer
respectively. This structure, called Two-Adlayer Asymmetric-Dimer
Model (TAADM) \cite{pollmann99,bernolc00,dangelo03}, was proposed
among several possible $3\times 2$ geometries  on the basis of
total-energy calculations including a detailed investigation of
the electronic structure and optical properties. In particular, in
this configuration the tilting of the upper dimers
(Si$_{D}-$Si$_{D}^{\prime}$) is responsible for the existence of a
gap between two surface bands, $\pi$-bonding (filled) and
$\pi^*$-antibonding (empty) on the Si$_{D}- $Si$_{D}^{\prime}$
dimers. The two second-layer dimers (Si$_{d}-$Si$_{d}^{\prime}$)
have a short length, and weaker silicon dimers are present at the
third layer (Si$_{b}-$Si$_{b}^{\prime}$). Let us call the latter
as {\sl Si atoms in the channel}.

H atoms can be bonded to the atoms of the external Si$_{D}-
$Si$_{D}^{\prime}$ dimers by weakening the strength of such bonds
and eliminating the buckling \cite{catellan,derycke01}. A further
supply of hydrogen could result in the di-hydrogenation of the Si
atoms of the external dimers or in the chemisorption of H on Si
atoms in the channel. Derycke \textit{et al.} \cite{souk}
suggested that only one Si atom of each channel dimer is bonded to
H, giving rise to a structure with a row of unsaturated Si DBs
(black sticks in Fig.~1b) in the [0$\bar{1}$1] direction at the
third surface layer.

To test this model, with the specific aim of interpreting the 
origin of metallicity in terms of surface bands, we
performed first-principle calculations based on the Density
Functional formalism. The surfaces were simulated by repeated
supercells. The unit supercell contained a slab with eight (or
sixteen \cite{layers}) atomic layers, and 13 \AA \thinspace\ of
vacuum to inhibit the interaction between neighboring replicas.
The bottom of the slab was terminated by a H layer saturating C
DBs, the top of the slab contained the extra adlayers of Si and H
atoms. All the investigated geometries were relaxed with respect
to both ionic and electronic degrees of freedom \cite{pwscf}, to
find the local minima starting from our initial conditions. The
atomic species were described by ultrasoft pseudopotentials
\cite{vanderbilt}, the electronic wavefunctions were expanded in
plane waves with a kinetic energy cutoff of 22 Ry. Brillouin Zone
(BZ) sums were performed at three special Monkhorst-Pack {\bf k}
points in the irreducible wedge of the two-dimensional (2D) BZ
\cite{note_BZ}. The PBE exchange-correlation functional was used
\cite{pbe}. We checked our results by performing test calculations
in the local spin density approximation.

We compared different patterns and amounts of hydrogenation of the
TAADM surface starting from different trial geometries. In the
following, the structures are labeled according to the H content
and adsorption layer: thus, \textsl{nH} indicates a hydrogenated
surface with \textsl{n} H atoms per unit cell adsorbed on top of
the outermost Si dimers of the $3\times 2$ reconstruction;
\textsl{nHmH} indicates a hydrogenated surface with \textsl{n} and
\textsl{m} H atoms per unit cell adsorbed on top and in the third
layer channel, respectively. (i) \textsl{2H} and \textsl{4H} are
characterized by the adsorption of H only on top of the outermost
Si dimers, with one and two H atoms for each top Si atom,
respectively; (ii) \textsl{3H} has two H atoms attached to
Si$_{D}$ and only one H atom attached to Si$_{D}^{\prime}$ on each
top dimer; (iii) \textsl{2H2H} includes H atoms at the outermost
Si dimers as in configuration \textsl{2H}, plus two H atoms per
cell at the Si$_{b}^{\prime}$ atoms in the channel, with one H
atom for each Si$_{b}-$Si$_{b}^{\prime}$ dimer (Fig.~1b,c) placed
at an initial H-Si$_{b}^{\prime}$ distance (structure
\textsl{2H2H$_u$} in Fig.~1b) typical of the H-Si single bond,
$\sim$1.5 \AA; (iv) \textsl{2H1H} (\textsl{3H1H}) can be obtained
from \textsl{2H2H} by removing (transferring to the outermost
dimers) half of the H atoms from the channel. In agreement with
previously published results \cite{catellan,derycke01}, the
\textsl{2H} and \textsl{4H} structures are semiconducting: The
outermost Si dimers are elongated from 2.30 \AA \thinspace\ to
2.42 \AA \thinspace\ in the \textsl{2H} geometry and are broken in
the \textsl{4H} one (Si-Si distance 3.16 \AA). A similar effect
occurs at  Si dimers of Si(100) \cite{h-si}. The electronic
structure in these cases exhibits surface bands that partially
overlap the upper and lower edges of the bulk gap. The occupied
and unoccupied bands remain well separated, preserving the
semiconducting behavior as in the case of the partial or full
hydrogenation of the Si(100) surface.

Assuming a uniform flux of atomic H, the adsorption of H atoms on
the Si atoms of the weak dimers in the channel is probable. This
gives rise to the most interesting case of the \textsl{2H2H}$_r$
structure. The starting configuration \textsl{2H2H}$_u$ for this
surface, suggested on experimental basis \cite{souk}, results
unstable: the H atoms bonded to Si$_{b}^{\prime}$ atoms (Fig. 1b)
move spontaneously to reach an equilibrium location bridging
Si$_{b}$ and Si$_{b}^{\prime}$ sites (\textsl{2H2H}$_r$, Fig. 1c). 
This migration was also obtained through extended calculations, 
with full atomic relaxation, performed with a larger supercell obtained 
by doubling the $3\times 2$ cell, to test that our results are not affected by the 
specific choice of periodic boundary conditions. 
In this migration, the H-Si$_{b}^{\prime}$ distance increases from
1.5 \AA~ to 1.68 \AA~ and becomes equal to the H-Si$_{b}$
distance. The Si$_{b}-$H$-$Si$_{b}^{\prime}$ angle becomes
133$^{\circ}$. In the relaxed geometry of Fig.~1c, the topmost
Si$_{D}-$Si$_{D}^{\prime}$ dimers are 2.36 \AA~ long (stretched by
5\% with respect to the clean TAADM surface) and flat, and the
H-Si$_{D}$ = H-Si$_{D}^{\prime}$ distance is 1.5 \AA. A full
account of the energetics for the hydrogenated SiC(100)-$(3\times
2)$ surface is beyond the scope of our work. However, let us point
out some considerations. Structures \textsl{2H}, \textsl{2H1H},
\textsl{2H2H}$_r$ differ from each other for the successive
addition of one H atom in the channel: therefore, they can be used
to estimate the energy gain upon subsurface hydrogenation. Our
results indicate that, in H-rich conditions, there is an energy
gain of 1.1 eV for the addition of one H atom per cell to the
structure \textsl{2H} to obtain \textsl{2H1H} 
(the total energies of two structures with a different H content are compared 
by adding to the energy of the H-deficient surface the local-spin-density 
free-atom chemical potential for H, consistently with the use of dissociated 
hydrogen in the experiments  \cite{souk}). A further energy gain of 0.7 eV is
realized by adding a second H atom per cell in the channel and
forming the equilibrium structure \textsl{2H2H}$_r$. The latter
configuration is unfavorable by 0.5 (0.28) eV/H-atom with respect
to the iso-stoichiometric configuration \textsl{4H}
(\textsl{3H1H}) where a higher H content is located at the
outermost dimers. The formation of a hydrogenated surface with the
same periodicity and H content in the various layers as in
\textsl{2H2H}$_r$, as revealed by Derycke and coworkers
\cite{souk}, is most likely induced by kinetic conditions attained
with a uniform exposure to atomic hydrogen \cite{nota}.

In agreement with the experiments, the \textsl{2H2H}$_r$ surface,
iso-stoichiometric with the claimed measured one \cite{souk},
exhibits a metallic character \cite{note_isostoich}. However, our
relaxed geometry is consistent with an alternative interpretation
for the electronic structure, not based on a DB band. Whereas the
H atoms adsorb on top of the first-layer Si dimers in a rather
conventional arrangement that realizes the saturation of surface
DBs, the fingerprint of this bizarre hydrogenated surface is the
presence of deep subsurface H adatoms. Additionally, the unusual
site selected by such adatoms is mainly responsible for the
modification of the surface bandstructure in the energy range of
the bulk bandgap. The H-Si$_{b}^{\prime}$ bonding scenario
\cite{souk} would imply saturation of a third-layer Si row
(Si$_{b}^{\prime}$) and depletion of the other (Si$_{b}$), thus
explaining the metallization in terms of a DB band. Instead, the
unusual bridge location for H induces a non-bonding
rehybridization of the third-layer Si dimers with a consequent
upward energy shift and surface-to-bulk charge transfer, as
detailed below.

\begin{figure}[tbp]
\includegraphics[scale=0.18]{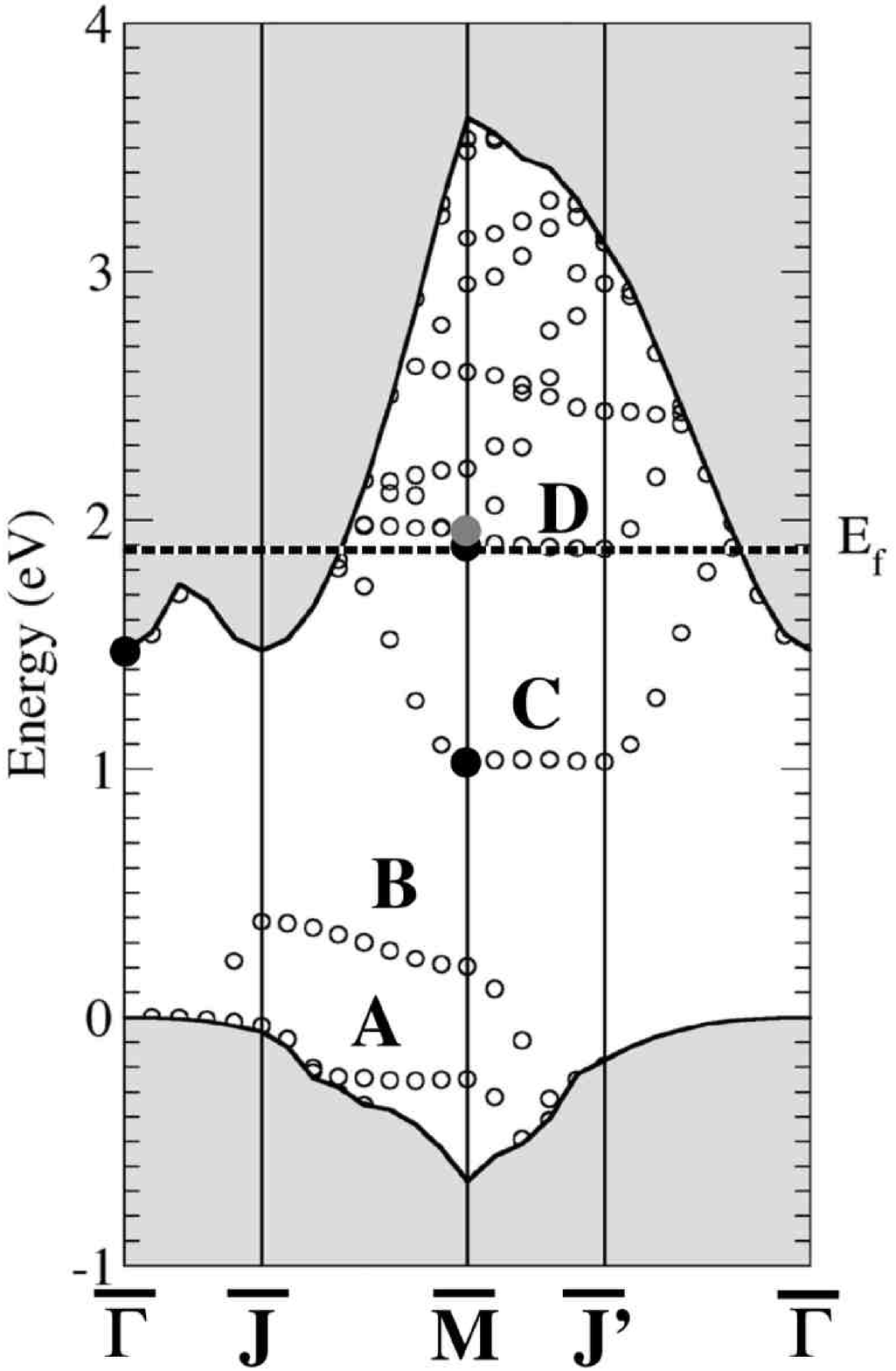} 
\caption{Computed bandstructure of the hydrogenated
\textsl{2H2H}$_r$ $\beta$-SiC(100)-3$\times$2 surface. Gray zones
mark the projected bulk bandstructure.}
\end{figure}
The  bandstructure of the \textsl{2H2H}$_r$ surface is shown in
Fig.~2. The bands are plotted along the edges of the irreducible
part of the $3\times 2$ 2D BZ. The shaded gray areas represent the
projected bulk bandstructure (PBBS). The 1.47~eV bandgap compares
reasonably well with the bulk DFT value of 1.38~eV (affected by
the renowned DFT underestimation, in this case 40\%). We find that
a few bands appear within the bandgap range, but they do not have
a DB character, as inferred from the wavefunction analysis discussed below. 
Two occupied surface bands A and B are present in
a small region of the 2D BZ slightly above the occupied PBBS
continuum. Separated from B by a small energy gap ($\simeq$
0.6~eV), other surface bands are revealed: They are partially
occupied. Band C lays in the upper region of the bulk bandgap. The
Fermi level is well inside the conduction band, and crosses also
other surface bands (D) in a small region of the 2D BZ. This is
consistent with the observation of a negative band bending at the
surface, as observed by STM \cite{souk}.

\begin{figure}[tbp]
\includegraphics[scale=0.56]{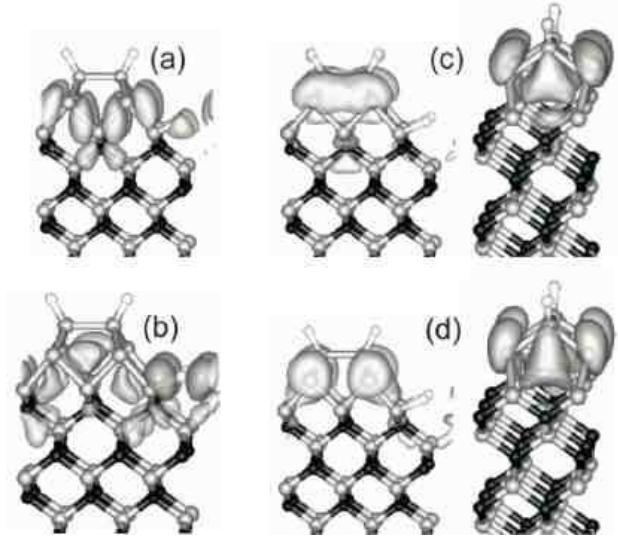} 
\caption{Isosurface plots of fully and partially occupied surface
states for the hydrogenated 3$\times$2 structure \textsl{2H2H}$_r$
(side views). Black, gray, white spheres represent C, Si, H atoms,
respectively. (a) Filled B band at $\bar{M}$. (b) Surface state at
$\bar{M}$ close to the D band, identified as a gray dot in Figure
2, almost at $E_f$. (c) C band at $\bar{M}$. (d) D band at
$\bar{M}$. In (c) and (d), the left (right) panels show views
along the [011] ([0$\bar{1}$1]) direction, slightly tilted.}
\end{figure}
The shape of the charge density distribution associated to the
most relevant electron states is shown in Fig.~3. Figure 3a
represents the band B at the $\bar{M}$ point: This is a state
localized at the subsurface H atoms ($s$-like shape) and at the
Si$_d$-Si$_b$ (Si$_{d}^{\prime}$-Si$_{b}^{\prime}$) backbonds.
Band A is given by similar orbitals. Bands A and B derive from a
couple of similar bands of the clean TAADM SiC(100)-$(3\times 2)$
surface, where the backbonds are instead hybridized with the
$\sigma$-bonding contribution of the channel Si dimers: They
remain just above the PBBS continuum. The metallic behavior of the
\textsl{2H2H}$_r$ surface stems both from the minimum of the
conduction band, which is occupied in the range of band bending,
and from the surface bands C and D. The charge density for the
state in Fig.~3b (explanation in the caption) clearly indicates a
non-bonding character between H and Si$_{b}^{\prime}$ (Si$_{b}$).
It is originated from one of the surface bands of the TAADM
structure with $\sigma$-bonding character on the channel
Si$_{b}$-Si$_{b}^{\prime}$ dimers: the dimer bonding
characteristic (clean surface) is transformed into the non-bonding
characteristic (hydrogenated surface), with a consequent upward
energy shift, upon subsurface H adsorption. In Figs. 3c and 3d we
show two other partially occupied surface states at $\bar{M}$,
belonging to bands C and D, respectively (black spots at $\bar{M}$
in Fig.~2). The dominant feature of the states in Figs.~3c and 3d
is the identical antibonding character along [0$\bar{1}$1] on
second-layer Si dimers (right panels), and a charge component also
on first-layer Si dimers. The difference between the two bands is
evident only along the [011] direction (left panels, bonding and
antibonding nature for Figs.~3c and 3d, respectively). Of the
three occupied surface bands that are present at the top of the
valence continuum in the clean surface \cite{pollmann99}, one due
to $\pi$-bonding states on the outermost dimers is shifted well
into the occupied bulk continuum, whereas the other two are split
into bands A and B that remain occupied, and C and D that are
partially occupied because, reaching energies degenerate with the
bottom conduction continuum, give origin to charge transfer to
bulk states. Summarizing the above description, our results
indicate that no metallic DB band localized on third-layer Si
atoms exists in the bandgap. The only electron states that can be
found within the bandgap and possess a charge component on the
third layer, have a clear non-bonding Si-H-Si character. The
strongest metallic contribution is due to the filling of the PBBS
continuum, and by the presence of partially filled surface states
which stem from the Si$_{b}$-Si$_{b}^{\prime}$ dimers (bonding)
and finally assume a non-bonding Si$_{b}$-H-Si$_{b}^{\prime}$
character. We note that, whereas on one hand the partial filling
of the bulk conduction band may be affected by the DFT
underestimation of energy gaps, the scenario denoted here suggests
a very common and well documented situation in surface/interface
science, characterized by the attainment of a two-dimensional
electron gas induced by band-bending effects \cite{band_bending}.

We estimate that of the extra 2 electrons due to third-layer
hydrogen, $\simeq$ 0.66 electrons are present in the upper surface
band C, while the remaining 1.33 electrons per 3$\times $2 cell
are in the bottom of the conduction band and in small portions of
the surface bands near the Fermi level (D). This geometry
(non-bonding bridge position) is rather unusual, yet compatible
with other investigations. In fact, a similar H coordination was
obtained for bulk SiC, in the case of interstitial hydrogen in the
neighborhood of a carbon vacancy, where  the H atom position is
equidistant from two neighboring Si atoms \cite{aradi01}. Our
calculations clearly indicate that the same ligand geometry is
assumed by the H atoms in the third layer at this unconventional
hydrogenated SiC(100)-$3\times 2$ surface, where no preferential
bond is created with one of the two neighboring Si$_b$ and
Si$_{b}^{\prime}$ atoms.

For completeness, in order to include in our investigation a truly
non-symmetric condition for sub-surface H adsorption, we have also
considered a geometry were the two H atoms are not in the channel
Si$_b$-Si$_{b}^{\prime}$, but at interstitial bulklike positions
inside the building blocks of the surface reconstruction. Such a
configuration, whose computed bandstructure reveals a
semiconducting character, is a local minimum of the total energy
surface: the energy is 0.42 eV per cell \cite{nota} higher than
that of the structure \textsl{2H2H}$_r$. 
We also computed the bandstructure for the asymmetric 
configuration \textsl{2H2H}$_u$ \cite{souk} without relaxing 
the ionic degrees of freedom: the outcome does not reveal a dangling-bond band in the gap. 

In conclusion, our first-principle calculations enable us to
support the metallic character of the hydrogenated
$\beta$-SiC(100)-(3$\times $2) surface \cite{souk}. We propose an
alternative explanation of the bandstructure, which is not
compatible with the formation of a DB band induced by
inequivalence of Si atoms at the broken third-layer Si dimers. (a)
The periodicity is preserved upon hydrogen chemisorption. (b) H
atoms chemisorb preferably on the outermost dimers and this does
not change the semiconductor character of the surface; the
separation between surface filled and empty bands increases. (c)
Although not energetically preferred, equilibrium structures with
H atoms in the subsurface channel can be found as local minima of
the potential energy surface. Third-layer H adsorption gives an
energy gain with respect to low-coverage regimes, but does not
result in additional covalent bonds. Instead, a channel H atom
chooses a \textit{bridge} position at the middle of the two
neighboring Si$_b$ and Si$_{b}^{\prime}$ atoms, depleting the
third-layer Si dimers of bonding charge. (d) These rehybridization
effects are translated into energy shifts well into the conduction
band of the orbitals originally localized at the channel Si
dimers, that simultaneously change their nature from
$\sigma$-bonding to non-bonding. (e) The Fermi level is in a
region where both surface and bulk bands are responsible for
conduction.

This work was funded by INFM through the Parallel Computing
Initiative and by MIUR-IT through FIRB-NOMADE.





\begin{thebibliography}{99}
\bibitem{layers} The energetics obtained from the calculations with a
8 atomic layers are in agreement with those obtained with 16
layers. The thicker slab describes correctly the relative energy
of the surface bands with respect to the projected bulk
bandstructure and is used for the computation and discussion of
the electronic properties in this work. The thinner slab is used
to evaluate the relative energy of the several computed
geometries.
\bibitem{nota} A full account of the reaction kinetics would require
the computation of several geometries to describe the potential
energy surface, and also the inclusion of zero-point energies in
the energetics estimates. This is beyond the scope of this
investigation, which is instead devoted to the electronic
properties.
\end{thebibliography}
\end{document}